# Superhydrophobic Auxetic Metamaterials

Glen McHale[1]*, Andrew Alderson[2], Steven Armstrong[1], Shruti Mandhani[2], Mahya Meyari[1], Gary G. Wells[1], Emma Carter[2], Rodrigo Ledesma-Aguilar[1], Ciro Semprebon[3], Kenneth E. Evans[4]

[1]Wetting, Interfacial Science & Engineering Laboratory, Institute for Multiscale Thermofluids, The University of Edinburgh, Edinburgh, EH9 3DF, Scotland UK.

[2]Materials & Engineering Research Institute, Sheffield Hallam University, Sheffield, S1 1WB, UK.

[3]Smart Materials & Surfaces Laboratory, Faculty of Engineering & Environment, Northumbria University, Newcastle upon Tyne, NE1 8ST, UK

[4]Department of Engineering, University of Exeter, Exeter, EX4 4QF, UK

*email: glen.mchale@ed.ac.uk

Superhydrophobic materials are often inspired by nature[1–4], whereas metamaterials are engineered to have properties not usually found in naturally occurring materials[5–8]. In both cases, the key that unlocks their unique properties is structure. Here, we show that a negative Poisson's ratio (auxetic) mechanical metamaterial[9–12] is capable of transforming into a unique type of superhydrophobic material. When stretched its surface has the counterintuitive property that it also expands in the orthogonal lateral direction. We model the change in the solid surface fraction as strain is applied and show it decreases as the space between solid elements of the auxetic lattice expands. This results in a unique dependence of the superhydrophobicity on strain. We construct experimental models illustrating the relationship between different states of strain and superhydrophobicity as the lattice structure transitions from an auxetic to a conventional (positive Poisson's ratio) one. The principles we have discovered offer a new approach to designing superhydrophobic materials for self-cleaning surfaces, droplet transportation, droplet encapsulation and oil-water separation.

Over recent years, shape and topography of surfaces have been a central focus in designing bespoke wetting properties into materials. By creating a bed-of-nails effect mimicking the Lotus leaf, it is possible to create superhydrophobic surfaces, which ball-up droplets far beyond the ca. 118° contact angle of Teflon™ possible with surface chemistry alone[1,2]. By switching the surface chemistry to hydrophilic, such a surface can be converted to a hemiwicking[13] or a superspreading surface[14]. Alternatively, by impregnating with a lubricating oil to replace the air in their lattice, it is possible to create a super-slippery liquid-infused porous surface (SLIPS) mimicking the *Nepenthes* pitcher plant, with virtually no resistance to droplet motion[15,16]. However, the prevailing paradigm in superhydrophobicity has been that the static arrangement of the lattice determines the solid surface fraction available to interact with a contacting liquid droplet and hence the wettability of the surface. There has been little attention to how fundamentally different arrangements of the lattice structure may be reconfigured dynamically, and the effect that such changes can have on the wettability of the surface itself.

In the simultaneously developing, but distinct, field of metamaterials, there has been a realization of the profound importance of lattice structure in determining unusual physical properties. In particular, auxetic mechanical metamaterials have the counterintuitive property that when they are stretched they expand in an orthogonal direction (Fig. 1a,b). Thus, unlike a normal material, an auxetic lattice can expand by the creation of additional space (in both the direction of stretch and orthogonal to that direction) between its solid components, which do not themselves stretch or compress. Since the balance of solid-to-air fraction at a surface controls extreme non-wetting and extreme wetting, auxetic materials would appear to be candidates for novel strain-controlled functional wetting materials. We therefore hypothesized that under tensile strain an auxetic lattice would necessarily reduce the solid fraction at the surface of a material and be capable of generating a transformation into a superhydrophobic surface.

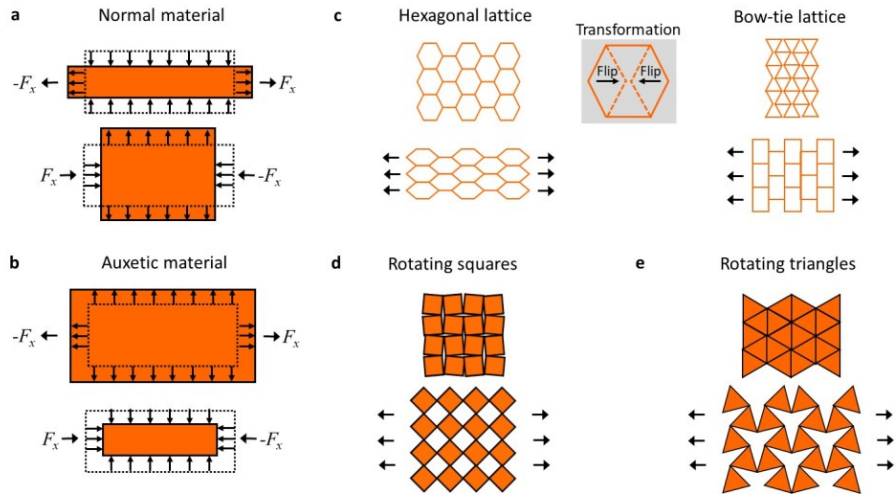

**Figure 1: Auxetic Metamaterial Surfaces.** Stretching and compression of **a**, a conventional material (positive Poisson's ratio) and **b**, an auxetic material (negative Poisson's ratio). **c**, Transformation of a conventional hexagonal lattice into an auxetic bow-tie lattice using rotation of rigid elements at nodal points[10,17]. Examples of rotating rigid shape-based auxetic materials. **d**, squares[12,18], and **e**, triangles[19]. In each of the auxetic cases, tensile strain induces a systemic decrease in the solid surface fraction in the auxetic lattice.

To develop our ideas, we first consider a conventional hexagonal lattice with inextensible solid elements able to rotate about their connecting nodes in response to a mechanical force applied to the lattice. When the lattice is stretched, the rotation of these solid elements at each lattice node causes the lattice to contract in the lateral direction[10] (Fig. 1c). This is what we intuitively expect from experience with materials such as rubber strips. This behavior is naturally characterized in terms of the Poisson's ratio, $v$, defined as the negative ratio of the transverse strain ($\varepsilon_{lateral}$) to the longitudinal strain ($\varepsilon_{axial}$), in the direction of the loading force, i.e. $v=-\varepsilon_{lateral}/\varepsilon_{axial}$. Positive and negative strains correspond to extension and contraction, respectively. For a conventional hexagonal lattice, the transverse and longitudinal strains have opposite signs and the Poisson's ratio is, therefore, positive.

We now imagine a simple transformation of the hexagonal lattice into a bow-tie lattice by flipping the angles at two opposing nodes from obtuse to acute in each hexagonal unit (Fig. 1c). When tensile strain is now applied to the lattice, the rotation of the solid elements at each lattice node causes a counterintuitive expansion of the lattice in the lateral direction[10,17] (Fig. 1c). Stretching a material or metamaterial causes an expansion of the surface area as long as $v < +1$ (see Supplementary Information). Importantly for its wettability, the increase in surface area of the bow-tie metamaterial is through an expansion of the space, and not the solid, within the lattice. As a consequence, the solid surface fraction, $f_S$, also decreases as the metamaterial is stretched when $v < +1$ (this includes the stretch-induced transition of the metamaterial from a bow-tie to an hexagonal lattice geometry). Stretching the metamaterial further, i.e., for $v > +1$, causes a decrease in surface area, with the corresponding increase in solid surface fraction. From the perspective of non-wettable materials, an initially hydrophobic auxetic bow-tie lattice metamaterial supporting a droplet in a Cassie-Baxter suspended state[20,21] will become systematically more hydrophobic and eventually superhydrophobic (including the minimum in solid surface fraction following the transition to a positive Poisson's ratio for an hexagonal lattice), before returning to a hydrophobic state upon further stretching. Similarly, it is possible to imagine new and unique superhydrophilic wettable, hemiwicking and liquid-infused auxetic materials.

We now illustrate how our concept for the wetting properties of auxetic metamaterials is applicable to a wide range of auxetic lattice structures and is not limited uniquely to bow-tie lattices. Another class of auxetic metamaterials uses tessellations of two-dimensional shapes connected at their corners so that each shape can rotate cooperatively about their corners. Thus, for example, a set of corner-connected rigid squares[12,18] as shown in fig. 1d rotate under strain into a diamond-shaped lattice with a Cassie solid surface fraction systematically decreasing from unity to 0.5. Other similar designs using, e.g. triangles[19] (Fig. 1e), also behave in an auxetic manner under strain. We therefore observe that there are many classes of auxetic metamaterials, whose surface wetting properties will change in a uniquely defined manner with strain.

We next model the solid surface fraction of a bow-tie auxetic surface with rotatable inextensible solid elements under tensile strain in the vertical ($x_2$) direction (Fig. 2a, Supplementary Information Fig. S6). The Poisson's ratio for this geometry has been derived previously to be $v_{21} = \sin\alpha(h_u/l_u + \sin\alpha)/\cos^2\alpha$ (geometrical parameters defined in the Supplementary Information), where the subscript refers to the loading ($x_2$) and transverse ($x_1$) directions. For simplicity, in the following we drop the subscript, i.e. $\nu \equiv \nu_{21}$. As the strain, $\varepsilon = \Delta L_{axial}/L_{axial}$, where $\Delta L_{axial}$ is the change in the axial length $L_{axial}$ (and, in this case, $L_{axial}$ corresponds to the unit cell length $X_2$ in the Supplementary Information), increases, the opposing arms of each bow-tie straighten (i.e. the negative value of α → 0°) and eventually an entirely rectangular lattice (α = 0°) is achieved. Beyond this value of strain, the angles at the two opposing nodes of each bow-tie shape invert to create a conventional hexagonal lattice (α > 0°). The solid surface fraction, $f_s = A_s/A$, where $A_s$ is the solid surface area and $A$ is the planar projection of total surface area, systematically decreases with strain until a minimum, which is always within the conventional lattice region, when $\nu = +1$, is reached (Fig. 2a; see Supplementary Information for a proof). The corresponding Cassie-Baxter contact angle, $\theta_{CB}$, of a droplet suspended on this geometry, is shown in fig. 2b (model parameters are detailed in the Supplementary Information). Here, $\theta_{CB}$ is defined through the weighted average $\cos\theta_{CB} = f_s\cos\theta_S + (1-f_s)\cos\theta_A$, where $\theta_S = 120°$ is the contact angle of a droplet on the solid and $\theta_A = 180°$ (i.e. $\cos\theta_A = -1$) represents the contact angle of a droplet on air. As strain is applied, the auxetic surface transforms from hydrophobic ($\theta_{CB} < 150°$) to superhydrophobic ($\theta_{CB} > 150°$) before becoming a conventional (non-auxetic), but still superhydrophobic, surface. Eventually, a maximum in superhydrophobicity is achieved before the lattice closes up and the material reverts to being a hydrophobic, but conventional surface. This reveals a unique property of the auxetic superhydrophobicity arising from the monotonic decrease in solid surface fraction with increasing strain. There is only ever a single value of strain for any one value of the Cassie-Baxter contact angle in the auxetic region. In contrast, the existence of a minimum solid surface fraction in the conventional region means a single value of Cassie-Baxter contact angle can correspond to either a strain above or a strain below that which characterizes the maximum.

To experimentally confirm the relation between lattice configurations at different states of strain and the wettability of the surface, we constructed a set of physical models (Supplementary Information). This enables direct comparison of the wetting response of the physical models with the predicted response of a honeycomb with rotatable inextensible solid elements under tensile strain, as opposed to straining a single physical model which would also be subject to flexing and stretching of the solid elements not accounted for in the predictive model used here. Our models used a polymer (SU-8) photo-lithographically patterned into bow-tie lattice micro-structures with solid surface fractions chosen to represent the full set of configurations across a strain curve (Fig. 3a). Each lattice was treated with a low-pinning hydrophobic coating, which gives advancing and receding contact angles of 105.2° and 102.5° on a flat control surface. By choosing the height of the walls (ca. 60 μm) to be larger than the typical void between the walls, penetration of water into the structure is inhibited, favoring a Cassie suspended droplet state. Figure 3b shows typical side profile images of droplets in contact with the surfaces of these structures. The full characterization of the wetting of these structures shows a systematic increase in the static contact angles, $\theta$, and, hence, hydrophobicity, as the lattices transition from bow-tie to rectangular, representing auxetic surface states (Fig. 3c).

Because the bow-tie lattice design is asymmetric, we report contact angles from two orthogonal viewing directions (Fig. 3c, inset), as well as their average. For large, positive $\alpha$ (e.g., $\alpha = 75°$ in Fig. 3a), the contact angle measured along orientation II is always smaller, suggesting that the droplet is able to spread more easily along the direction of the solid elements. Notably, this property is reversed as the geometry crosses over to the auxetic region. A maximum corresponding to an extreme non-wetting contact angle of ca. 150º with a droplet in a suspended state is observed in the conventional (hexagonal) lattice region before falling rapidly as the lattice closes up and the solid surface fraction increases. The two continuous curves show the predicted hydrophobicity for these lattice designs using the Cassie-Baxter model with the advancing and receding contact angles of 105.2º (upper curve) and 102.5º, respectively, and show a remarkable agreement between the model and the experiments.

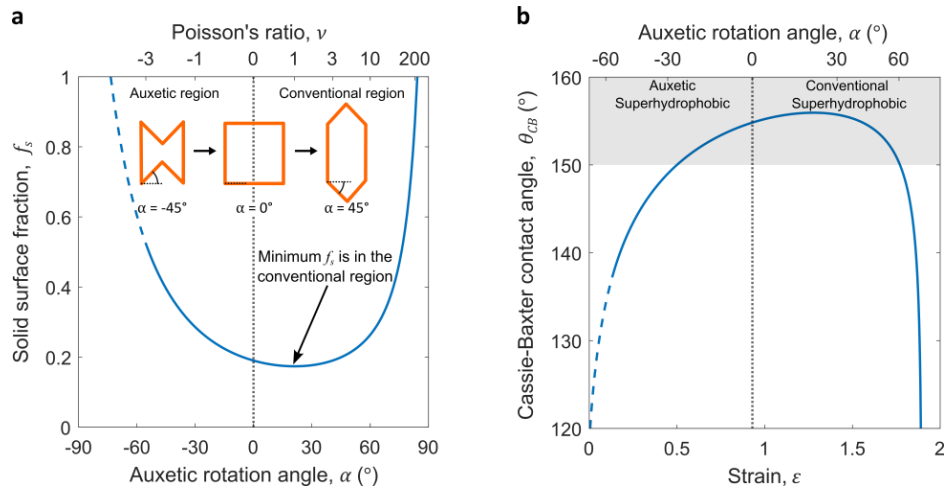

**Figure 2: Model Strain Dependence of a Hydrophobic Auxetic Surface. a,** Predicted changes of solid surface fraction, $f_s$, with auxetic rotation angle, $\alpha$. The initially auxetic bow-tie lattice is transformed into a conventional hexagonal lattice for positive auxetic rotation angle, as indicated by the Poisson's ratio, $\nu$ (top axis). The minimum solid surface fraction occurs in the conventional lattice region when $\nu = +1$. **b,** Predicted Cassie-Baxter contact angle, $\theta_{CB}$, for suspended state droplets with applied tensile strain, $\varepsilon$. Strain-induced superhydrophobicity occurs for both the auxetic and conventional regions, but there is a unique value of contact angle for each value of strain in the auxetic case. The zero of strain has been defined to correspond to a closed auxetic lattice with the maximum sold surface fraction, $f_s$. Model parameters are detailed in the Supplementary Information.

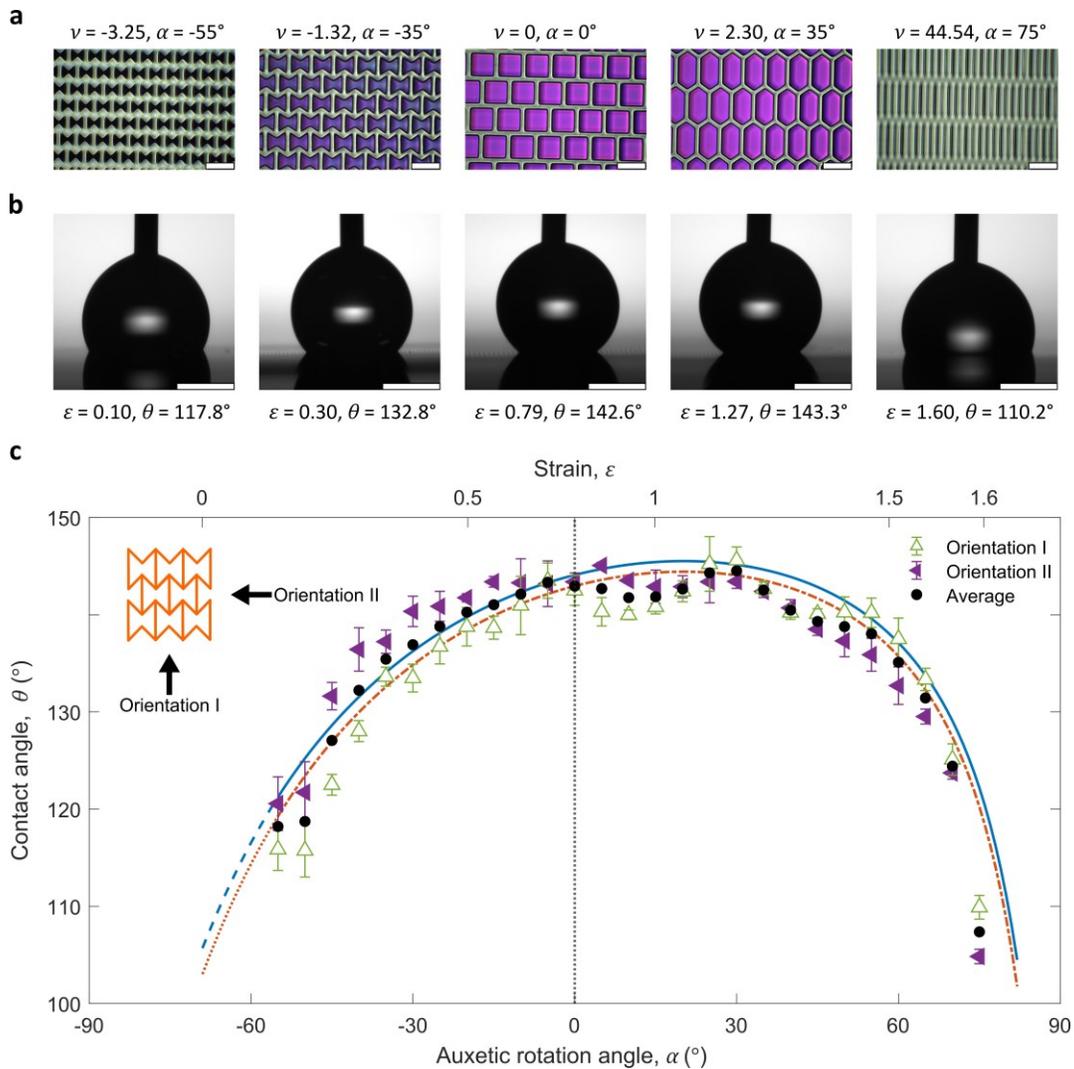

**Figure 3: Model of a Hydrophobic Auxetic Surface. a**, Top images of low-pinning hydrophobic surfaces representing different strained states of a bow-tie lattice (scale bar 100 μm). **b**, Side profile images of water droplets in contact with the surfaces (scale bar 500 μm). **c**, Data for contact angle viewed from two orthogonal directions (Inset: Orientation I and II). Continuous curves show model predictions using the lattice design parameters and the advancing and receding contact angles on fluorosilanized SU-8 surfaces (upper and lower curves, respectively). The dashed and dotted portions of the curves indicate values of the angle for which lattice solid elements overlap.

Next, we developed a process to create micro-patterned polydimethylsiloxane (PDMS) membranes with auxetic surface wetting properties (Supplementary Information). A layer of a photoresist (SPR220-7) was spin-coated to create a film on a silicon wafer and then patterned to a depth of ca. 20 μm using exposure to ultraviolet (UV) light. PDMS was then spin coated across the pattern to fill the pattern and cured. Subsequently, the excess above the pattern was etched away to expose the lattice ribs. A final soak in acetone then dissolved the photoresist releasing the auxetic PDMS membranes. These membranes were mounted in a micro-stretching apparatus and observed from above to confirm auxetic behavior (Fig. 4a and Supplementary

Information: Video 1). Droplets in contact with these membranes tended to penetrate into the void spaces between membrane ribs even when hydrophobized with fluorosilane (Fig. 4b). This could be prevented using a superhydrophobic nanoparticle coating (Fig. 4c), but doing so created a completely non-adhesive surface onto which a droplet could not be detached onto the membrane from the syringe (Supplementary Information: Video 2). When droplets were deposited onto such superhydrophobic membrane, the contact angle was insensitive to the level of strain applied (Supplementary Information Fig. S5). Since the rigidity of a thin sheet scales with the cube of its thickness, these membranes have little rigidity normal to their surface, and elastocapillary effects[22,23] are observed with the droplet bending the membranes out of plane (Fig. 4b). Such an effect could offer the opportunity to use the synclastic (double) curvature[11,17,24] of an auxetic membrane to wrap a droplet surface[22] without causing wrinkling or creasing.

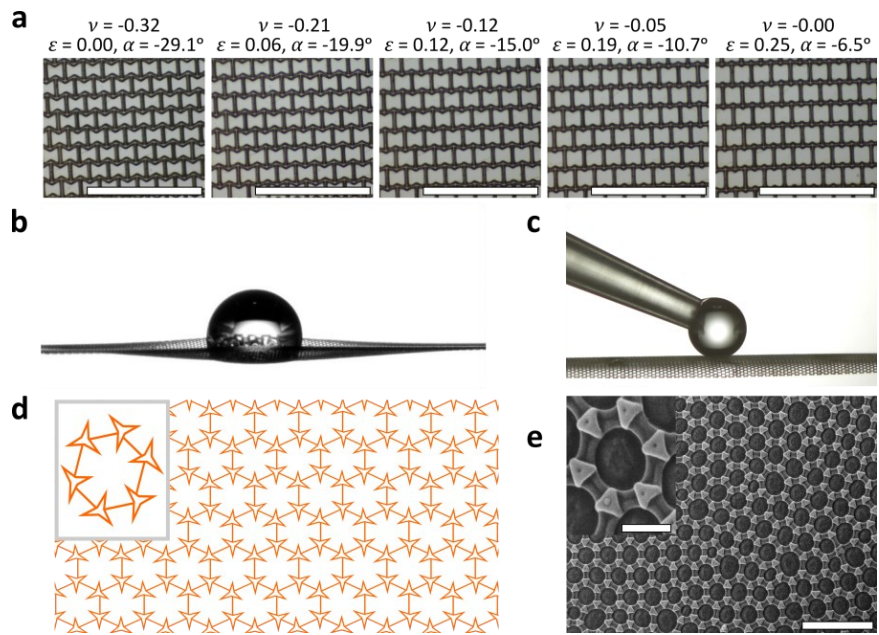

**Figure 4: Auxetic Membranes. a**, Top images of an auxetic fluorosilanized polydimethyl siloxane (PDMS) membrane in different states of strain with corresponding auxetic rotation angle and Poisson's ratio values (scale bar 500 µm). Side profile images of water droplets in contact with the surface of **b**, an auxetic fluorosilanized PDMS membrane, and **c**, an auxetic PDMS membrane possessing a superhydrophobic nanoparticle coating. The membrane in **b** is distorted out of plane by elastocapillary forces due to the strength of the surface tension forces from the droplet. **d**, an auxetic lattice pattern which resembles the structure observed on **e**, the cuticle water repelling surface of the soil-dwelling springtail *Orchesella cincta*[25] (SEM images provided by A. E. Filippov; scale bars in the main image and the detail are 1 µm and 200 nm, respectively).

Finally, we return to the contrasting ideas that metamaterials are materials with properties that are uncommon, but that superhydrophobic surfaces are themselves inspired by nature. Natural biological systems grow and their surfaces have to adapt both to the change in curvature of their surfaces and to the environment in which they live, which is often water or a partially wet or water-clogged material. A common adaptation of insects to breathing underwater is the use of breathing holes with a surface-attached layer of air (a "plastron")[26] whose air-water interface acts as an oxygen-carbon dioxide exchange membrane. The plastron is a naturally occurring feature when a superhydrophobic material is immersed in water. In the case of insects such as *Orchesella cincta*[25], a type of springtail (Collembola) that lives in water-logged ground, the plastron is achieved with hydrophobic surface features which prevent the breathing holes being flooded. It is therefore interesting to contrast images of the cuticle surface of the Springtail with a classic connected star-type auxetic design[27] (Fig. 4d,e). It is possible that the springtail exoskeleton has adapted such that as the underlying surface stretches the star-shaped surface feature spacing expands in an auxetic manner or alternatively an auxetic expansion is achieved by the straightening of the edges on each star-shaped surface feature. Although studies of this geometry are beyond our current report, we believe these observations should motivate a wider search for naturally occurring auxetic wetting properties in biological systems.

The use of metamaterial concepts in the design and control of wetting is a hitherto unrecognized opportunity for new types of material surfaces with unique properties. In this work, we have used auxetic principles to create strain-controlled hydrophobicity and superhydrophobicity exemplified by lattice models and membranes. These principles can be extended to other types of auxetic lattices. They will also apply to the wetting of the surfaces of structure independent auxetic materials with unknown geometries, but known elastic properties, such as shape memory foams, provided they have small pores, such that interfacial tension forces dominate. We have also suggested that the uncommon properties of metamaterial wettability might have natural examples. The design principles presented here provide a foundation for new types of surfaces relevant to super-water repellent applications, liquid encapsulation and micro-reactors, and may also have application to strain controllable liquid-liquid separations.

1.  Neinhuis, C. & Barthlott, W. Characterization and distribution of water-repellent, self-cleaning plant surfaces. *Ann. Bot.* **79**, 667–677 (1997).

2.  Barthlott, W. & Neinhuis, C. Purity of the sacred lotus, or escape from contamination in biological surfaces. *Planta* **202**, 1–8 (1997).

3.  Sun, T., Feng, L., Gao, X. & Jiang, L. Bioinspired surfaces with special wettability. *Acc. Chem. Res.* **38**, 644–652 (2005).

4.  Bhushan, B. & Jung, Y. C. Natural and biomimetic artificial surfaces for superhydrophobicity, self-cleaning, low adhesion, and drag reduction. *Prog. Mater. Sci.* **56**, 1–108 (2011).

5.  Pendry, J. B. Negative refraction makes a perfect lens. *Phys. Rev. Lett.* **85**, 3966–3969 (2000).

6.  Nicolaou, Z. G. & Motter, A. E. Mechanical metamaterials with negative compressibility transitions. *Nat. Mater.* **11**, 608–613 (2012).

7. Kadic, M., Bückmann, T., Schittny, R. & Wegener, M. Metamaterials beyond electromagnetism. *Reports Prog. Phys.* **76**, 126501 (2013).

8. Hewage, T. A. M., Alderson, K. L., Alderson, A. & Scarpa, F. Double-negative mechanical metamaterials displaying simultaneous negative stiffness and negative Poisson's ratio properties. *Adv. Mater.* **28**, 10323–10332 (2016).

9. Evans, K. E., Nkansah, M. A., Hutchinson, I. J. & Rogers, S. C. Molecular network design. *Nature* **353**, 124–124 (1991).

10. Evans, K. E. & Alderson, A. Auxetic materials: Functional materials and structures from lateral thinking! *Adv. Mater.* **12**, 617–628 (2000).

11. Lakes, R. S. Foam Structures with a Negative Poisson's Ratio. *Science (80-. ).* **235**, 1038–1040 (1987).

12. Lakes, R. S. Negative-Poisson's-Ratio Materials: Auxetic Solids. *Annu. Rev. Mater. Res.* **47**, 63–81 (2017).

13. Bico, J., Tordeux, C. & Quéré, D. Rough wetting. *Europhys. Lett.* **55**, 214–220 (2001).

14. McHale, G., Shirtcliffe, N. J., Aqil, S., Perry, C. C. & Newton, M. I. Topography driven spreading. *Phys. Rev. Lett.* **93**, 036102 (2004).

15. Wong, T.-S. *et al.* Bioinspired self-repairing slippery surfaces with pressure-stable omniphobicity. *Nature* **477**, 443–447 (2011).

16. Lafuma, A. & Quéré, D. Slippery pre-suffused surfaces. *Europhys. Lett.* **96**, 56001 (2011).

17. Alderson, A. & Alderson, K. L. Auxetic materials. *Proc. Inst. Mech. Eng. Part G J. Aerosp. Eng.* **221**, 565–575 (2007).

18. Grima, J. N. & Evans, K. E. Auxetic behavior from rotating squares. *J. Mater. Sci. Lett.* **19**, 1563–1565 (2000).

19. Grima, J. N., Jackson, R., Alderson, A. & Evans, K. E. Do zeolites have negative Poisson's ratios? *Adv. Mater.* **12**, 1912–1918 (2000).

20. Cassie, A. B. D. & Baxter, S. Wettability of porous surfaces. *Trans. Faraday Soc.* **40**, 546 (1944).

21. Quéré, D. Wetting and Roughness. *Annu. Rev. Mater. Res.* **38**, 71–99 (2008).

22. Py, C. *et al.* Capillary Origami: Spontaneous Wrapping of a Droplet with an Elastic Sheet. *Phys. Rev. Lett.* **98**, 156103 (2007).

23. Bico, J., Reyssat, É. & Roman, B. Elastocapillarity: When surface tension deforms elastic solids. *Annu. Rev. Fluid Mech.* **50**, 629–659 (2018).

24. Evans, K. E. The design of doubly curved sandwich panels with honeycomb

cores. *Compos. Struct.* **17**, 95–111 (1991).

25. Filippov, A. E., Kovalev, A. & Gorb, S. N. Numerical simulation of the pattern formation of the springtail cuticle nanostructures. *J. R. Soc. Interface* **15**, 20180217 (2018).

26. Crisp, D. J. & Thorpe, W. H. The water-protecting properties of insect hairs. *Discuss. Faraday Soc.* **3**, 210 (1948).

27. Grima, J. N., Gatt, R., Alderson, A. & Evans, K. E. On the potential of connected stars as auxetic systems. *Mol. Simul.* **31**, 925–935 (2005).

**Supplementary Information** is available in the on-line version of this paper.

**Acknowledgements** The authors were supported in this work by funding from the UK Engineering & Physical Sciences Research Council (EP/T025158/1 and EP/T025190/1). M.M. was supported by the EPSRC CDT in Soft Matter for Formulation and Industrial Innovation, EP/S023631/1. We would like to thank A.E. Filippov for kindly providing the SEM images of figure 4e.

**Author Contributions** G.M. and A.A. conceived the project. GM, A.A., G.G.W., R.L-.A. and E.C. supervised the research. G.M., G.G.W. and S.A. designed the materials fabrication methods. G.G.W. and S.A. designed the measurement systems. S.A. fabricated the samples. M.M. carried out the surface treatment of the samples. S.A. and M.M. carried out wettability measurements and analysis. G.M., A.A. and S.M. developed the modelling with assistance from R.L.-A., E.C. and C.S. G.M., A.A., K. E., G.G.W., R.L.-A., E.C and C.S. discussed models, their interpretation and applicability to different systems. G.M. and R.L.-A. wrote the manuscript with inputs from all other authors.

**Author Information** The authors declare no competing financial interests. Readers are welcome to comment on the online version of the paper. Correspondence and requests for materials should be addressed to G.M. (glen.mchale@ed.ac.uk).

# Superhydrophobic Auxetic Metamaterials

Glen McHale[1]*, Andrew Alderson[2], Steven Armstrong[1], Shruti Mandhani[2], Mahya Meyari[1], Gary G. Wells[1], Emma Carter[2], Rodrigo Ledesma-Aguilar[1], Ciro Semprebon[3], Kenneth E. Evans[4]

[1]Wetting, Interfacial Science & Engineering Laboratory, Institute for Multiscale Thermofluids, The University of Edinburgh, Edinburgh, EH9 3DF, Scotland UK;

[2]Materials & Engineering Research Institute, Sheffield Hallam University, Sheffield, S1 1WB, UK.

[3]Smart Materials & Surfaces Laboratory, Faculty of Engineering & Environment, Northumbria University, Newcastle upon Tyne, NE1 8ST, UK

[4]College of Engineering, Mathematics and Physical Sciences, University of Exeter, Exeter, EX4 4QF, UK

*email: glen.mchale@ed.ac.uk

## EXPERIMENTAL METHODS

### SU-8 Based Physical Model for an Auxetic Bow-Tie Lattice.

To make the physical model the manufacturers data sheet for SU8-3035 permanent negative resist (Kayaku Advanced Materials, Inc.) was used as a basis for the process and modified for specific equipment. In a clean room environment, 3" silicon wafers are coated with a 500±10 nm oxide layer *via* plasma enhanced chemical vapor deposition STS Multiplex PECVD) to promote adhesion between the wafer and photoresist (Fig. S1a). To further enhance adhesion the wafer is then treated with hexamethyldisilazane (HMDS) for 10 minutes by keeping the wafer in a closed container with an open vial containing 2-3 drops of HMDS. The photoresist (MEGAPOSIT™ SU8-3035, Kayaku Advanced Materials) is then spin coated onto the oxide coated silicon wafer: 500 rpm for 10 seconds with 100 rpm s$^{-1}$ acceleration; 2000 rpm for 30 seconds with 300 rpm s$^{-1}$; then soft baked for 10 minutes at 95°C (Fig. S1b). The thickness of photo resist is measured as 62±1 μm using a stylus profilometer (DektakXT, Bruker). The photoresist coated wafer is then patterned via a direct-write photolithography machine (MicroWrite ML3 Pro, Durham Magneto Optics Ltd) exposing the desired pattern to UV light (5000mJ/cm$^2$) (Fig. S1c). The exposed photoresist coated wafer is then post-exposure baked for 2 minutes at 95°C (Fig. S1d). After post-exposure bake the un-reacted photoresist is removed by developing the wafer via submersion in propylene glycol methyl ether acetate (PGMEA) for 2 minutes (Fig. S1e), then rinsed with fresh PGMEA for a few seconds, followed by rinsing with IPA and dried with compressed nitrogen. The lattices were designed with parameters $h$ = 100 μm, $l$ = 50 μm and $t$ = 10 μm. Once produced, the lattices were characterized using the software imageJ. Five measurements of $h$, $l$ and $t$ were taken for each auxetic rotation angle, $α$, between 25° to 50° to give the average geometrical parameters of the sample as $h$ = 103.8 ± 1.5 μm, $l$ = 48.8 ± 1.0 μm and $t$ = 13.9 ± 1.1 μm. The measured values were then used in the analytical model to predict the Cassie angle as a function of the auxetic rotation angle.

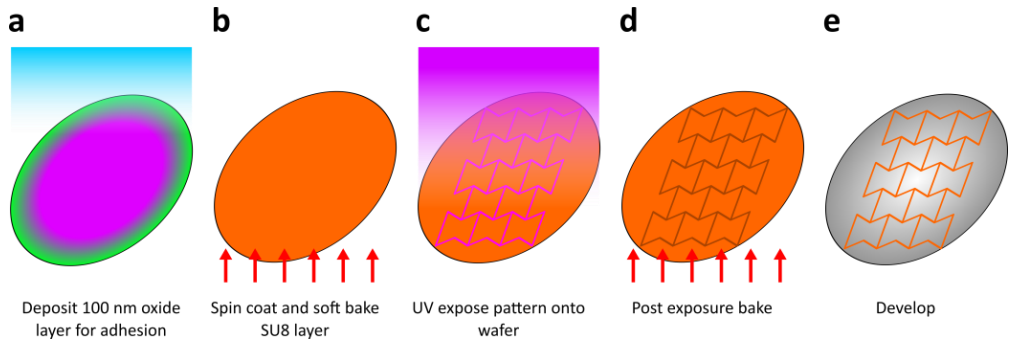

**Supplementary Figure S1. Fabrication Process of SU-8 Based Physical Model for an Auxetic Bow-Tie Lattice. a**, Silicon wafer with deposition of 100 nm oxide layer for improved adhesion. **b**, Spin coat and soft bake SU8 layer. **c**, UV expose pattern onto resist using direct-write photolithography. **d**, Post exposure bake. **e**, Develop to leave only auxetic bow-tie physical model pattern.

## Fabrication of Auxetic Bow-Tie Lattice Membranes.

Typically, micro-patterning of PDMS is achieved by creating a 'stamp' out of SU-8 photoresist[1]. This usually creates a micron-scale pattern on the top surface of a millimetre-scale bulk of PDMS. The bulk PDMS aids the lift-off from the SU-8 mould. To observe auxetic behaviour, the pattern needs to go through the entire material, not just the top surface, this adds an additional challenge in releasing the PDMS from the mould. Dissolving the mould allows release of the PDMS mesh without damaging the complex shapes of the auxetic network. SPR220-7 is a positive photoresist which is dissolvable in acetone, making it a good candidate for a dissolvable mould. SPR220-7 typically allows the creation on ~7 μm thick films. In order to make thicker auxetic networks a method of layering successive thin films of SPR220-7 was achieved by following a modified methodology adapted from Koukharenko *et al.*[2]

First the wafer is treated with hexamethyldisilazane (HMDS) for 10 minutes by keeping the wafer in a closed container with an open vial containing 2-3 drops of HMDS. Once treated, SPR220-7 is poured onto the wafer in a spin coater covering approximately 50% of the wafer. The wafer is then spin coated for: 2 min at 350 rpm with 100 rpm s$^{-1}$ acceleration; then 20 sec at 1000 rpm with 100 rpm s$^{-1}$ acceleration. This is then placed on a hotplate for 1 min at 90°C. A second layer is applied with the same spin-coater and hotplate parameters. The wafer is then placed in an oven (Tannay Jr) for 55 min at 90 C. For the third and final coat, once again the same spin-coater and hotplate parameters are used. This time the wafer is placed in the oven for 90 min at 90C (Fig. S2a).

The photoresist coated wafer is then patterned *via* direct-write photolithography (MicroWrite ML3 Pro, Durham Magneto Optics Ltd) exposing the desired pattern to UV light, 358 nm wavelength UV light at 1800 mJ/cm$^3$ exposure dose. Instead of a typical post exposure bake, the wafer is left for 24 hours after exposure to allow the UV light to break down the photoresist. Early development tests showed a post exposure bake would cause unwanted bubbles in the exposed areas, damaging the features of the mould. The wafer is developed using MF-24A with 15 minutes development time (Fig. S2b).

PDMS is prepared a 5:1 ratio of polymer to cross-linker to improve rigidity over standard 10:1 ratios[3]. The PDMS is placed in a vacuum desiccator for 30 min before use to remove any bubbles created mixing the polymer and cross-linker. The PDMS is poured onto the wafer covering approximately 50% of the wafer, then spun for 1 min at 6000 rpm. This is then cured in the oven for 60 min at 90C (Fig. S2c).

To ensure the PDMS rests inside the mould without laying over the top of the features, the wafer is briefly placed in a RIE (JLS Etcher) 15 minutes at 200 W RF power with 60 sccm $CF_4$ and 15 sccm $O_2$. This etches the PDMS down, removing any excess on top of the mould (Fig. S2d). After etching the PDMS can be removed from the mould by submersion into acetone (Fig. S2e). The membranes are quickly removed to avoid swelling of the PDMS in the acetone and placed into DI water, here they rest between the water/air interface where they can be removed and placed into the strain apparatus using a 3D printed membrane handling tool (Fig. S3a).

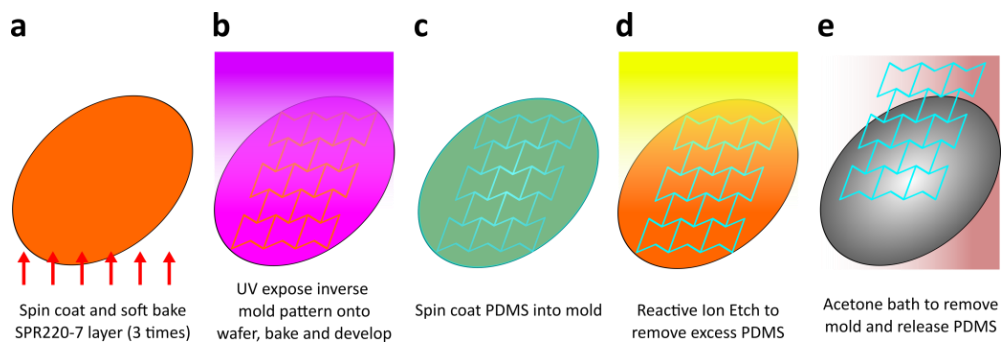

**Supplementary Figure S2. Fabrication Process of Auxetic Bow-Tie Lattice Membranes. a**, Spin coat 3 successive layers of SPR 220-7 to create a thick ~50 um layer. **b**, UV expose inverse bow-tie pattern to create a negative mold. **c**, Spin coat PDMS into mold. **d**, Shed excess PDMS using reactive ion etching. **e**, Release PDMS membrane by dissolving SPR220-7 mold in acetone.

## Hydrophobic Coating of Membranes and SU8-Based Physical Model.

To achieve low contact-line pinning and hydrophobic properties, samples were coated with a polymer brush Slippery Omiphobic Covalently Attached Liquid (SOCAL) layer[4]. First, a thin glass-like layer was added to the surface by spin coating (750 rpm for 40 s) sigmacote (Sigma Aldrich) onto the SU8 structured surface. The sigmacote was left to dry overnight before applying the SOCAL layer. SOCAL was then applied using the optimised method developed by Armstrong *et al.* [5] The wafer was treated with an air plasma at 30% power, 20 minutes and 15 sccm (Henniker HPT-200) to add OH radicals to the glass surface. The wafer was then dipped into a reactive solution of IPA, dimethyldimethoxysilane (Sigma Aldrich), and sulphuric acid (Fisher Chemical)(90, 9, and 1%wt.) for 10 s, and then slowly withdrawn. The wafer is then placed in a bespoke humidity-controlled environment at 60% Relative Humidity and room temperature (18 – 22 C) for 20 minutes. During this step, an acid-catalyzed graft polycondensation of dimethyldimethoxysilane creates a homogeneous layer of PDMS chains grafted to the surface. Any unreacted solution is then rinsed off with IPA, Toluene and DI water to complete the process.

To make PDMS membranes superhydrophobic, a commercial spray coating is used (GLACO Mirror Coat, Nippon Shine). PDMS membranes held on a 3D printed membrane handling tool are sprayed from 20 cm distance. Excess liquid on the handling tool is removed via blotting with a Whatman lens tissue to prevent the membrane from migrating along the tool. The membrane is left to air dry for 30 minutes. This coating process is then repeated 3 times to ensure the coating is sufficient to make the surface superhydrophobic.

**Strain Measurements on Membranes.**

A bespoke strain apparatus was created using two motorized linear stages (MT1-Z8, Thorlabs) controlled *via* software to adjust displacement of the motor position. To load the membrane, the 3D printed tool (Fig. S3a) is attached across the linear stages with additional 3d printed clamps over the top of the membranes (Fig. S3b). The membrane secured to either motorized stage. The supporting centre of the 3D printed tool is cut at the point they join the linear stages to allow the membrane to be suspended (Fig. S3c). Finally the stages are moved apart until the membrane is no longer sagging but with care to not induce a strain (Fig. S3d). This displacement of the linear stages is taken as zero strain. Further displacement of the stage positions inducing a strain in the membrane. Cameras are positioned to the side and underneath of the membrane to capture side view shadowgraph images of droplets on the membranes under varying strain and capture a bottom view of the membrane lattice changes under strain (Fig. S3 b-d). The images of the membrane are analyzed using open-source software Image-J, measuring deformation in the lattice structure as a function of strain. A 5x5 network of unit cells in the centre of the membrane was selected. The membrane was filmed under strain and 5 frames were selected at random, with the first one being at zero strain. The length of the network of unit cells along both vertical and horizontal axes for each frame was measured. Total strain along both these directions was calculated using $\varepsilon_{i\_total} = \frac{l_i - l_i^0}{l_i^0}$ where $\varepsilon_{i\_total}$ is the total strain along direction $i$, $l_i$ is length of the network along $i$, and $l_i^0$ is length of the network at zero strain along $i$. True strain was then calculated using $\varepsilon_{i\_true} = \ln\left(\frac{l_i}{l_i^0}\right)$ and $\varepsilon_{1\_true}$ (i.e. true transverse strain) was plotted against $\varepsilon_{2\_true}$ (i.e. true loading strain) to determine instantaneous Poisson's ratio $\nu_{21}$ for each image from the slope of the graph.

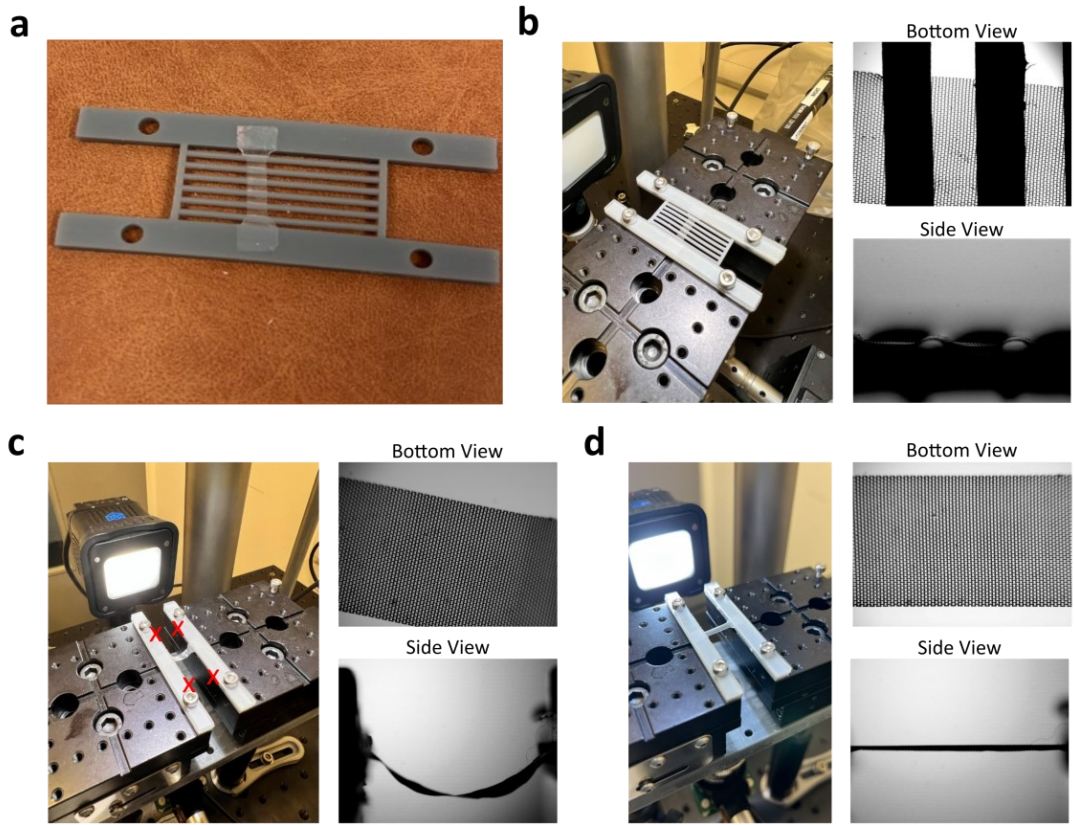

**Supplementary Figure S3. Loading of Strain Measurement Apparatus. a**, PDMS auxetic bow-tie lattice membrane held on bespoke 3D printed membrane handling tool. **b**, Membrane handling tool attached across two motorized linear stages. **c**, Center support clipped out of membrane handling tool at points denoted by red "X" symbols. **d**, Suspended membrane held at zero strain without sagging.

## Contact Angle Measurements.

Side profile images of droplets are taken using shadowgraphy technique on a bespoke goniometer to measure static contact-angle and contact-angle hysteresis of the SU8-based physical model surface. A Microfluidic syringe pump (Exigo, Cellix) is held vertically on a high load vertical stage (VAP10/M, Thorlabs)(Fig S4a) above a triple-axis xyz micron-adjustment stage (PT3/M, Thorlabs.)(Fig S4d). This allows precise dosing and placement of droplets onto the surface. A camera and a small-aperture spotlight (Lumecube 2.0, Lumecube) are placed either side of the droplet to capture images and video sequences (Fig. S4b,c) respectively.

For static contact-angle measurements on the SU8 based physical model, 1.0 ± µL de-ionised water (18.2 MΩ·cm ultrapure type 1) is dispensed and suspended from a flat tipped needle (32 gauge, 0.23 ± 0.01 mm). The suspended droplet is gently lowered to the surface until it attaches to the surface. An image of the droplet profile is taken shortly after touch down on the surface. The static-contact angle is then measured using a 3rd degree polynomial fit of the tangent between the contour of the droplet and the base diameter from the image using open-source droplet shape analysis software (pyDSA).[6]

For static contact-angle measurements on the flurosilanised PDMS auxetic bow-tie lattice membranes at different strains 1.0 ± 0.02 µL de-ionised water (18.2 MΩ·cm ultrapure type 1) is dispensed using a variable volume manual pipette. This is gently lowered onto the surface by hand. An image of the droplet is then taken immediately after touchdown and analyzed using pyDSA as above. 10 droplets are placed at each strain and to give an average static contact-angle for each engineering strain (Fig. S5).

For contact-angle hysteresis measurements, a video is recorded of a droplet dosing/aspiration sequence at 20 frames per second. A 4.0 µL droplet is placed on the surface; the droplet is allowed to relax for 20 seconds to an equilibrium state; with the needle still placed in the top of the droplet 2.0 µL is then dosed at 6 µL s$^{-1}$ allowing another 20 seconds before the next step in the sequence; 6 µL is then aspirated from the droplet at -6 µL s$^{-1}$; care is taken during the aspiration to not withdraw so much liquid that the droplet is removed from the surfaces before a receding contact-angle is observed. As previous, pyDSA is used to analyze the droplet contact-angle, throughout the video sequence. The advancing angle is determined as the contact-angle the droplet makes with the surface instantly before the contact-line begins to move during the dosing procedure. The receding angle is determined as the contact-angle the droplet makes with the surface instantly before the contact-line begins to move during the aspiration procedure. The contact-angle hysteresis is then calculated as difference between the advancing and receding angle.

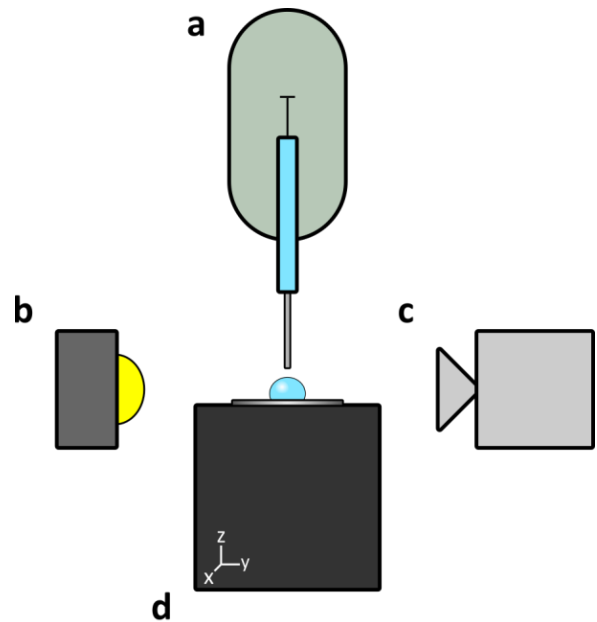

**Supplementary Figure S4. Contact Angle Measurement Apparatus.** a, Programmable Syringe pump suspended over experiment on a high load vertical stage. b, Small aperture backlight. c, Macro zoom lens and camera. d, Triple axis stage.

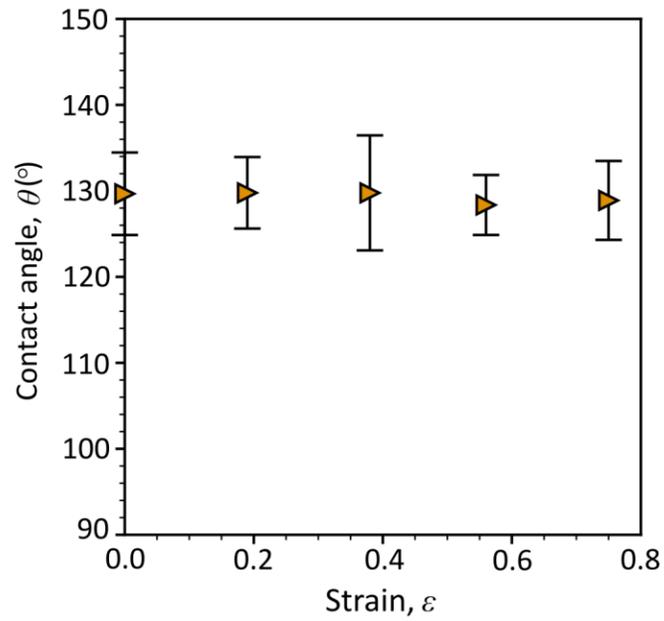

**Supplementary Figure S5. Static Contact Angles on Auxetic Bow-Tie Lattice Membranes.** Static contact-angle of DI water on flurosilanised PDMS auxetic bow-tie lattice membranes at different engineering strains. Images of the membrane at different levels of strain are shown in Fig 4a of the main manuscript.

# THEORETICAL ANALYSIS

## Model for the Superhydrophobicity of an Auxetic Bow-Tie Lattice.

Supplementary Fig. S6 shows a unit cell of a conventional honeycomb, represented by the dashed red rectangle. Analytical expressions for the projected lengths $X_1$ and $X_2$ of the unit cell are developed by representing these distances with respect to the variable parameter, honeycomb angle, $\alpha$.

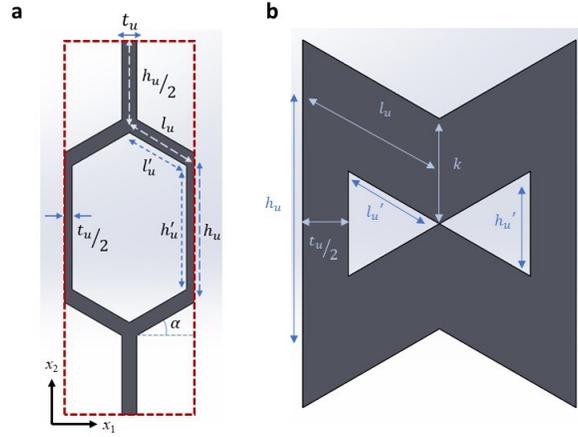

**Supplementary Figure S6. Geometry of a Conventional Honeycomb. a**, Schematic of the parameterization of the unit cell. Negative values of the rotation angle, $\alpha$, indicate a clockwise rotation giving an auxetic lattice and positive values indicate an anticlockwise rotation giving a conventional honeycomb. **b**, Maximum thickness of the diagonal rib, $l_u'$ when considering all other parameters

The length of the unit cell along $x_1$ is,

$$X_1 = 2\, l_u \cos \alpha \tag{S-e1}$$

and the length of the unit cell along $x_2$ is,

$$X_2 = 2\,(h_u + l_u \sin \alpha) \tag{S-e2}$$

The area of the unit cell, $A_{\text{cell}} = X_1 X_2$, is

$$A_{\text{cell}} = 4 l_u \cos \alpha\, (h_u + l_u \sin \alpha) \tag{S-e3}$$

Positive $\alpha$ corresponds to an anti-clockwise angle with respect to the horizontal axis, $x_1$, for the rib shown in the conventional honeycomb (Fig. S6a), and a clockwise, and thus negative, angle for re-entrant (auxetic) honeycomb geometries. Thus, the rotation angle $\alpha$ determines whether the unit cell is auxetic (positive Poisson's ratio) or conventional (negative Poisson's ratio) and can parameterize the evolution between the two types (Fig. S7).

| Parameter | Symbol | Value |
|---|---|---|
| Honeycomb angle | $\alpha$ | -55° to 84° |
| Mesh depth | $d_u$ | 5 units |
| Rib thickness | $t_u$ | 1 unit |
| Rib height | $h_u$ | 10 units |
| Rib length | $l_u$ | 5 units |

**Supplementary Table 1. Table of Standard Honeycomb Parameters.**

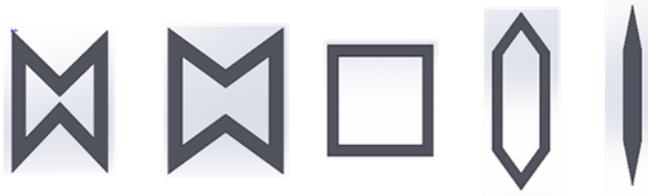

**Supplementary Figure S7. Example Evolution of a Unit Cell from Auxetic to Conventional.** The evolution is parameterized by the rotation angle $\alpha$ with auxetic shapes corresponding to -90°<$\alpha$<0 and conventional shapes corresponding to 0≤$\alpha$<90°. The transition from auxetic to conventional occurs when $\alpha$=0° which corresponds to a rectangular unit cell.

Strain can be calculated using $\Delta X_i / X_i$ where $\Delta X_i$ is the difference between the final and initial length of unit cell, and $X_i$ is the initial length, and $i$ represents the direction (either 1 or 2).

The limits for the honeycomb angle are determined by the rib thickness ($t_u$), height ($h_u$), and length ($l_u$). The inner vertical rib height, $h'_u$, and inner rib length, $l'_u$, are dependent on $t_u$, the rib thickness, and honeycomb angle $\alpha$,

$$h'_u = h_u - \frac{t_u(1-\sin\alpha)}{\cos\alpha} \tag{S-e4}$$

$$l'_u = l_u - \frac{t_u}{2\cos\alpha} \tag{S-e5}$$

The Cassie-Baxter state is when a droplet remains in a suspended state, bridging between the gaps (or pores) on the surface of a material. It remains in contact with the solid surface area. A simplifying assumption is to ignore any meniscus in the pore area and assume the liquid-vapor interface remains flat.

The pore area within the unit cell in Fig. 6a is given by,

$$A_{\text{pore}} = 4l'_u \cos\alpha \, (h'_u + l'_u \sin\alpha) \tag{S-e6}$$

From eq. S-e4 and eq. S-e5,

$$A_{\text{pore}} = 4 \cos \alpha \left(l_u - \frac{t_u}{2\cos\alpha}\right)\left[\left(h_u - \frac{t_u(1-\sin\alpha)}{\cos\alpha}\right) + \left(l_u - \frac{t_u}{2\cos\alpha}\right)\sin\alpha\right] \quad \text{(S-e7)}$$

The pore surface fraction, is the ratio of the pore area to the unit-cell area, and is 1 minus the Cassie solid surface fraction, $f_s$,

$$1 - f_s = \frac{\left(l_u - \frac{t_u}{2\cos\alpha}\right)\left[\left(h_u - \frac{t_u(1-\sin\alpha)}{\cos\alpha}\right) + \left(l_u - \frac{t_u}{2\cos\alpha}\right)\sin\alpha\right]}{l_u(h_u + l_u \sin\alpha)} \quad \text{(S-e8)}$$

The Cassie solid surface fraction is the ratio of solid area to the unit-cell area,

$$f_s = 1 - \frac{\left(l_u - \frac{t_u}{2\cos\alpha}\right)\left[\left(h_u - \frac{t_u(1-\sin\alpha)}{\cos\alpha}\right) + \left(l_u - \frac{t_u}{2\cos\alpha}\right)\sin\alpha\right]}{l_u(h_u + l_u \sin\alpha)} \quad \text{(S-e9)}$$

The Cassie-Baxter contact angle, $\theta_{CB}$, can be calculated from the weighted average on the solid and air fractions[7],

$$\cos\theta_{CB} = f_s \cos\theta_S - (1 - f_s) \quad \text{(S-e10)}$$

where $\theta_S$ is the contact angle of a droplet on the material of solid and $\cos\theta_A = -1$ because $\theta_A = 180°$ for air.

Although the honeycomb angle may vary from −90° to 90°, the added thickness, $t_u$, imposes limits on those extremes. The Cassie solid surface fraction, $f_s$, starts close to unity and decreases as the diagonal ribs of the honeycomb rotate and cause expansion in the system, creating more pore area. With increasing strain, the auxetic honeycomb opens up reaching a honeycomb angle value of 0°, resembling a rectangle. Under continued strain, the system enters a positive Poisson's ratio regime and starts closing transversely again as the honeycomb ribs rotate and now shorten the length $X_1$ instead of expanding. Once $\alpha$ reaches its maximum limit, the system reaches a Cassie solid surface fraction close to 1 and is (almost) a complete solid surface.

The effect of rib thickness, $t_u$, on the Cassie-Baxter angle can be studied by varying it from its minimum thickness of 0 units to its maximum thickness when opposing apexes meet and divide the pore in a unit cell into two distinct parts (Fig. 6b). The maximum thickness of the diagonal rib, $l_u$, is given by,

$$k = \frac{t_{l_u\_max}}{\cos\alpha} \quad \text{(S-e11)}$$

and

$$k = h_u + 2l_u \sin\alpha \quad \text{(S-e12)}$$

which implies,

$$t_{l_u\_max} = \cos\alpha\,(h_u + 2l_u \sin\alpha) \quad \text{(S-e13)}$$

The maximum thickness of vertical rib is given by,

$$t_{h_u\_max} = 2\, l_u \cos \alpha \tag{S-e14}$$

Resultant values from eq. S-e12 and eq. S-e13 are compared, and the lowest of the two represent maximum rib thickness, $t_u$, in honeycomb geometries.

The range of thickness that can be achieved in honeycombs is higher for conventional systems when compared to auxetic ones as the re-entrant structure restricts the maximum allowable thickness. This means that conventional honeycombs can achieve a Cassie solid surface fraction of ~1, i.e. a complete solid, whereas this is not possible for auxetic honeycombs. For a given rib thickness and magnitude of honeycomb angle, $|\alpha|$, Cassie Baxter contact angles are always greater in conventional honeycombs than auxetics.

For example, assuming as 2:1 ratio of rib height to rib length of a honeycomb with a solid material contact angle of $\theta_S$ =105° and rib thickness, $t_u$=2.5 units, a honeycomb angle of $\alpha$=-30° results in a Cassie Baxter contact angle of $\theta_{CB}$=120°, however its conventional counterpart of $\alpha$=+30° honeycomb angle results in a Cassie-Baxter angle of $\theta_{CB}$=140°. The Cassie-Baxter solid surface fraction for conventional honeycombs is always less than that of its auxetic counterpart, i.e. honeycombs with an angle of $\alpha$=30°/45°/60° have less solid-to-unit-cell-area compared to auxetic honeycombs with angles of $\alpha$=-30°/-45°/-60°, respectively. The higher amounts of solid in auxetic honeycombs thus results in lower Cassie-Baxter contact angles.

**Theoretical Analysis of Surface area vs Poisson's ratio**

Consider a rectangular material of dimensions $X$ and $Y$, and surface area

$$A = XY. \tag{S-e15}$$

We now consider an infinitesimal strain $\varepsilon_y$ (applicable to both linear and non-linear responses) along the $y$ direction:

$$d\varepsilon_y = \frac{dY}{Y}. \tag{S-e15}$$

The corresponding infinitesimal strain along $x$ direction, $\varepsilon_x$, obeys

$$d\varepsilon_x = \frac{dX}{X}. \tag{S-e16}$$

Hence, the Poisson's ratio, $\nu_{yx}$, is

$$\nu_{yx} = -\frac{d\varepsilon_x}{d\varepsilon_y}, \tag{S-e17}$$

which yields

$$d\varepsilon_x = -\nu_{yx} d\varepsilon_y. \tag{S-e18}$$

Let us now consider the change in surface area with displacement along $y$, i.e.,

$$\frac{dA}{dY} = X + Y\frac{dX}{dY} = X\left(1 + \frac{Y}{X}\frac{dX}{dY}\right) = X\left(1 + \frac{d\varepsilon_x}{d\varepsilon_y}\right) = X(1 - \nu_{yx}). \qquad \text{(S-e19)}$$

For positive (extension) change in $Y$, the surface area increases when $\nu_{yx} < +1$, and decreases when $\nu_{yx} > +1$.

The rate of change of surface area with strain is given by

$$\frac{dA}{dY}Y = \frac{dA}{d\varepsilon_y} = XY(1 - \nu_{yx}) = A(1 - \nu_{yx}). \qquad \text{(S-e20)}$$

For positive (extension) change in strain along $y$, the rate of change of surface area with strain increases as $\nu_{yx}$ decreases when $\nu_{yx} < +1$, and increases as $\nu_{yx}$ increases when $\nu_{yx} > +1$. There is no change in surface area with strain when $\nu_{yx} = +1$.

For a lattice structure with constant solid surface area ($A_S$) under strain (e.g. a hinging honeycomb), the solid surface fraction, $f_S = A_S/A$, also does not change with strain when $\nu_{yx} = +1$, corresponding to a minimum in the solid fraction vs strain curve.

Note that the range of Poisson's ratio for *isotropic* materials is $-1 < \nu < +0.5$. Hence the decrease in surface area under tension noted above when $\nu_{yx} > +1$ is not possible for isotropic materials, and the minimum in the solid fraction vs strain curve is not observed.

Similarly, the enhanced rate of change of surface area with strain at high magnitude of (negative or positive) Poisson's ratio requires the material to be *anisotropic*.

**Video Auxetic Bow-tie Membrane Stretching**

> **Supplementary Video SV1. Auxetic Bow-tie Membrane Stretching.** Video demonstration of auxetic bow-tie membrane going from unstrained to strained states with unit cell expansion along strain and perpendicular to the strain. Filmed at 10 fps, playback at 30 fps. Horizontal ribs of unit cell are 100 μm for an indication of scale.

**Video of Attempted deposition of Droplet onto a Superhydrophobic Coated Membrane**

> **Supplementary Video SV2. Attempted Deposition of Droplet onto Superhydrophobic Coated Membrane.** Video demonstration of difficulty to place a 1.0 μL droplet onto superhydrophobic GLACO coated auxetic bow-tie membrane. Filmed and playback at 100 fps. Pipette tip is 0.90 ± 0.05 mm for an indication of scale.

**Supplementary References**

1. Bubendorfer, A., Liu, X. & Ellis, A. V. Microfabrication of PDMS microchannels using SU-8/PMMA moldings and their sealing to polystyrene substrates. *Smart Mater. Struct.* **16**, 367–371 (2007).

2.  Koukharenko, E., Kraft, M., Ensell, G. J. & Hollinshead, N. A comparative study of different thick photoresists for MEMS applications. *J. Mater. Sci. Mater. Electron.* **16**, 741–747 (2005).

3.  Seghir, R. & Arscott, S. Extended PDMS stiffness range for flexible systems. *Sensors Actuators, A Phys.* **230**, 33–39 (2015).

4.  Wang, L. & McCarthy, T. J. Covalently Attached Liquids: Instant Omniphobic Surfaces with Unprecedented Repellency. *Angew. Chemie Int. Ed.* **55**, 244–248 (2016).

5.  Armstrong, S., McHale, G., Ledesma-Aguilar, R. & Wells, G. G. Pinning-Free Evaporation of Sessile Droplets of Water from Solid Surfaces. *Langmuir* **35**, 2989–2996 (2019).

6.  Launay, G. PyDSA: Drop Shape Analysis in Python. *https://framagit.org/gabylaunay/pyDSA_gui* https://framagit.org/gabylaunay/pyDSA_gui (2018).

7.  Shirtcliffe, N. J., McHale, G., Atherton, S. & Newton, M. I. An introduction to superhydrophobicity. *Adv. Colloid Interface Sci.* **161**, 124–138 (2010).